\documentclass[useAMS,usenatbib]{mn2e}

\usepackage{graphicx}
\usepackage{epstopdf}

\usepackage{dcolumn}
\usepackage{bm}
\usepackage{amsmath}
\usepackage{mathrsfs}
\usepackage{amsfonts}
\usepackage[usenames]{color}

\usepackage{epstopdf}

\newcommand{\be} {\begin{equation}}
\newcommand{\ee} {\end{equation}}
\newcommand{\de} {\,\mathrm{d}}
\newcommand{\bea}{\begin{eqnarray}}
\newcommand{\eea}{\end{eqnarray}}
\newcommand{\bdm}{\begin{displaymath}}
\newcommand{\edm}{\end{displaymath}}
\newcommand{\ba} {\begin{array}}
\newcommand{\ea} {\end{array}}

\newcommand{\bfg}  {\begin{figure}}
\newcommand{\efg}  {\end{figure}}
\newcommand{\incgr} {\includegraphics}
\newcommand{\mbf}{\mathbf}

\newcommand{\bit}{\begin{itemize}}
\newcommand{\eit}{\end{itemize}}

\newcommand{\fnl}{f_{\rm NL} }

\title{Constraints on Primordial Non-Gaussianity from a Needlet Analysis of the WMAP-5 Data}
\author[D.~Pietrobon et al.]
{Davide Pietrobon$^{1,2}$
\thanks{E-mail:davide.pietrobon@roma2.infn.it},
 Paolo Cabella$^{1,4}$,
 Amedeo Balbi$^{1,3}$,
 Giancarlo de Gasperis$^1$,
\newauthor and Nicola Vittorio$^1$ \\
\\
$1$ Dipartimento di Fisica, Universit\`a di Roma ``Tor Vergata'', via della Ricerca Scientifica 1, 00133 Roma, Italy \\
$2$ Institute of Cosmology and Gravitation, University of Portsmouth, Mercantile House, Portsmouth PO1 2EG, United Kingdom \\
$3$ INFN Sezione di Roma ``Tor Vergata'', via della Ricerca
Scientifica 1, 00133 Roma, Italy\\
$4$ Dipartimento di Fisica, Universit\`a La Sapienza, P.~le A.~Moro 2, Roma, Italy\\ }

\begin{document}

\maketitle

\label{firstpage}

\begin{abstract}
We look for a non-Gaussian signal in the WMAP 5-year temperature
anisotropy maps by performing a needlet-based data analysis. We use
the foreground-reduced maps obtained by the WMAP team through the
optimal combination of the W, V and Q channels, and perform realistic
non-Gaussian simulations in order to constrain the non-linear coupling
parameter $\fnl$. We apply a third-order estimator of the needlet
coefficients skewness and compute the $\chi^2$ statistics of its
distribution. We obtain $-80<\fnl<120$ at 95\% confidence level, which
is consistent with a Gaussian distribution and comparable to previous
constraints on the non-linear coupling. We then develop an estimator
of $\fnl$ based on the same simulations and we find consistent
constraints on primordial non-Gaussianity. We finally compute
the three point correlation function in needlet space: the constraints
on $\fnl$ improve to $-50<\fnl<110$ at 95\% confidence level.
\end{abstract}


\section{Introduction}
With the increasing amount of high-quality observations performed in
the last decade
(\cite{Hinshaw2008WMAP5,Acbar2008,CBI2007,Quad2008Wu,Quad2008Pryke,Quad2008Hinderks,Masi2006,Maxipol2007Johnson}),
statistical tests of the CMB temperature anisotropy pattern are
getting more and more accurate.  This has made possible to test one of
the basic tenets of the standard cosmological scenario, i.e.\ that the
primordial density perturbations follow a Gaussian distribution. This
is a definite prediction of the simplest inflationary models
\citep{Guth1981,Sato1981,Linde1982,AlbrechtSteinhardt1982}: the detection of primordial deviations from
Gaussianity would be a smoking gun for more complicated
implementations of the inflationary mechanism, such as those of multi-fields 
\citep{LythWands2002,LindeMukhanov2006,AlabidiLyth2006}, ekpyrotic \citep{Mizuno2008,Khoury2002PhDT} or cyclic scenarios \citep{SteinhardtTurok2002,LehnersSteinhardt2008}.

When dealing with the search for a non-Gaussian statistics in real
data, two major issues have to be addressed. One has to do with the
statistical tools used to analyse the data and detect deviations from
Gaussianity: not only can these deviations be very subtle and elusive,
but they could be generated by processes that are not directly related
to the primordial density perturbations --- such as unremoved
astrophysical foregrounds \citep{Cooray2008WL,Serra2008} or instrumental systematics. The other issue
is theoretical, and relates to the way the non-Gaussianity is parameterised: while there is only one way to realize a Gaussian distribution, non-Gaussian statistics can be produced in countless ways. One then has to assume a non-Gaussian parameterisation which relates in some sensible way to an underlying early universe scenario.    

The latter issue is usually addressed by introducing a parameter
$\fnl$, which quantifies the amplitude of non-Gaussianity as a
quadratic deviation with respect to the primordial Gaussian
gravitational potential $\Phi_{\rm L}$, i.e.:
\begin{equation}
\Phi(x)=\Phi_{\rm L}(x)+\fnl \left[ \Phi^2_{\rm L}(x) -\langle \Phi_{\rm L}^2(x)\rangle \right]
\end{equation}
The major advantage of this parameterisation is that, regardless of the specific underlying early universe model, it can represent the second-order approximation of any non linear deviation from Gaussianity. For an excellent review on this topic see \cite{Bartolo2004NGreview}.

From the point of view of data analysis, a number of techniques have
been proposed in the past few years to quantify the level of deviation
from Gaussian statistics in the data.
The most used one in harmonic space is the bispectrum \citep{Luo1994,Heavens1998,SpergelGoldberg1999,KomatsuSpergel2001,Cabella2006IntBis}. 
The bispectrum is defined as the three-point correlation function, and an estimate of $\fnl$ through the bispectrum requires the sum over all the triangle
configurations. Since this is extremely time-consuming, regardless whether
the computation is performed in harmonic space or pixel
space, \cite{KomatsuSpergelWandelt2005} have proposed a fast cubic
estimator based on the Wiener filter matching, which reduces
considerably the computational challenge. This estimator has been
further developed by \cite{CreminelliEtAl2005} introducing a linear correction 
which allows the correct treatment of the anisotropic noise, and
finally optimised \citep{YadavKomatsuWandelt2007} and extended to polarisation
measurements \citep{YadavEtAl2007}.
Recently, \cite{YadavWandelt2007} applied the cubic estimator to the
WMAP 3-year data, finding a detection of a primordial non-Gaussian signal at more than 2.5 sigma. An indication of a primordial non-Gaussian signal has been also found by the WMAP collaboration in the analysis of the 5-year
dataset, although with a lower confidence level  \citep{WMAP5Komatsu2008}. An interesting discussion on optimal and sub-optimal estimators can be found in \cite{SmithZaldarriaga2006}. Se for further details \cite{YuLu2008} and \cite{Babich2005}.

Concerning the methods in pixel space, \cite{DeTroia2007B2K,Acbar2008,Hikage2006MinkFunc,Curto2008Archeops,Curto2007Archeops}
applied Minkowski functionals to several CMB datasets;
\cite{Cabella2005LocCurv} applied local curvature on WMAP 1-year data
and \cite{Monteserin2006} developed scalar statistics using the
Laplacian as a tool to test Gaussianity. Alternative indicators based on skewness and
kurtosis have been proposed by \cite{Bernui2008}.
Tests based on wavelets were
applied to WMAP 1-year data \citep{Vielva2004,Mukherjee2004}, and WMAP
5-year data by \cite{Curto2008waveNG} and
\cite{McEwen2008}. Wavelets have been also used in CMB studies
\citep{Martinez-G2002} to identify anomalies in WMAP data
\citep{McEwen2008,Wiaux2008Morph,Cruz2008,Cruz2007ColdSpotW3,Vielva2007ColdSpot,Wiaux2006,Cruz2005ColdSpot,Vielva2004,Pietrobon2008AISO},
denoising \citep{Sanz1999Denoising}, point sources extraction
\citep{Cayon2000PointS,Gonzalez-Nuevo2006PointS}. Very recently
\cite{Vielva2008} constrained primordial non-Gaussianity by means of
N-point probability distribution functions.

In this work, we constrain the primordial non-Gaussianity by applying for the first time a
novel rendition of spherical wavelets called {\em needlets} to the
WMAP 5-year CMB data. The needlet construction is discussed by
\cite{NarcowichPetrushevWard2006,Baldi2006,Baldi2007,Marinucci2008NEE,Geller2008Mat}.
Needlets have a number of interesting properties which make them a
promising tool for CMB data analysis: they live on the sphere, without
relying on any tangent plane approximation, are
quasi-exponentially localised in pixel space and can be easily
constructed from a filter function with a finite support in harmonic
space. Combined with the specific shape of this function, this guarantees a tiny level of
correlation among needlets, which are only marginally affected by the
presence of sky cuts. An exhaustive
discussion can be found in \cite{Pietrobon2006ISW} and \cite{Marinucci2008NEE}. Recently, the
needlet formalism has been extended to the the polarisation field, as
discussed by \cite{Geller2008SpinNee}. Here we try to test whether needlets can lead to interesting constraints on the non-Gaussian amplitude, when compared to previous estimates of $\fnl$.

This paper is organised as follows: in Sec.~\ref{sec:need_form} we
review the needlets formalism; in Sec.~\ref{sec:sims_res} we
describe the statistical analysis of 
WMAP 5-year data and the constraints on $\fnl$; we summarise our conclusions  in Sec.~\ref{sec:concl}.

\section{Needlets Formalism}
\label{sec:need_form}
We perform our analysis of the non-Gaussianity of WMAP 5-year data by means of needlets. So far needlets have been successfully applied to the study of the CMB in the context of the detection of the Integrated Sachs-Wolfe effect (ISW) \citep{Pietrobon2006ISW}, the power spectrum estimation \citep{Fay2008PS} and the study of deviations from statistical isotropy \citep{Pietrobon2008AISO}. We refer to the work by \cite{Marinucci2008NEE} for details on the construction of a needlet frame and a detailed analysis of its statistical properties. Here we remind that a set of needlets has one free parameter,  $B$,  which controls the width of the filter function in harmonic space. The filter function is non-vanishing only within a certain multipole range, thus allowing to keep trace of the angular information of the signal; the specific shape of the filter function also ensures a sharp localisation in pixel space. The correlation among needlets of the same set characterised by different $j$ is much smaller than any other wavelet-like frame found in literature so far and can be easily described analytically. These features are particularly useful when looking for weak signals, like the non-Gaussianity one, which can be hidden by sky cuts and anisotropic noise.

Formally, a needlet, $\psi_{jk}$, is a quadratic combination of spherical harmonics which looks like
\be
  \psi _{jk}(\hat\gamma) = \sqrt{\lambda_{jk}}\sum_{\ell}b\Big(\frac{\ell}{B^{j}}\Big)\sum_{m=-\ell}^{\ell}\overline{Y}
_{\ell m}(\hat\gamma)Y_{\ell m}(\xi _{jk}).
  \label{eq:needlets_expansion}
\ee
where $\hat\gamma$ is a generic direction in the sky and
$\xi_{jk}$ refers to a cubature point on the sphere for the $j$
resolution. The function $b(\ell/B^j)$ is the filter in $\ell$-space and
$\lambda_{jk}$ is a cubature weight number, which, following
\cite{Marinucci2008NEE}, has been set for practical purpose equal to
$1/N_p$, being $N_p$ the number of pixels in the Healpix
\citep{Gorski2005Healpix} scheme for the resolution chosen.
We show an example of the filter function for one set among those used in the analysis in Fig.~\ref{fig:b_filter}. In the following we use the short notation $b_\ell$ for the filter function.
\bfg
  \incgr[width=\columnwidth]{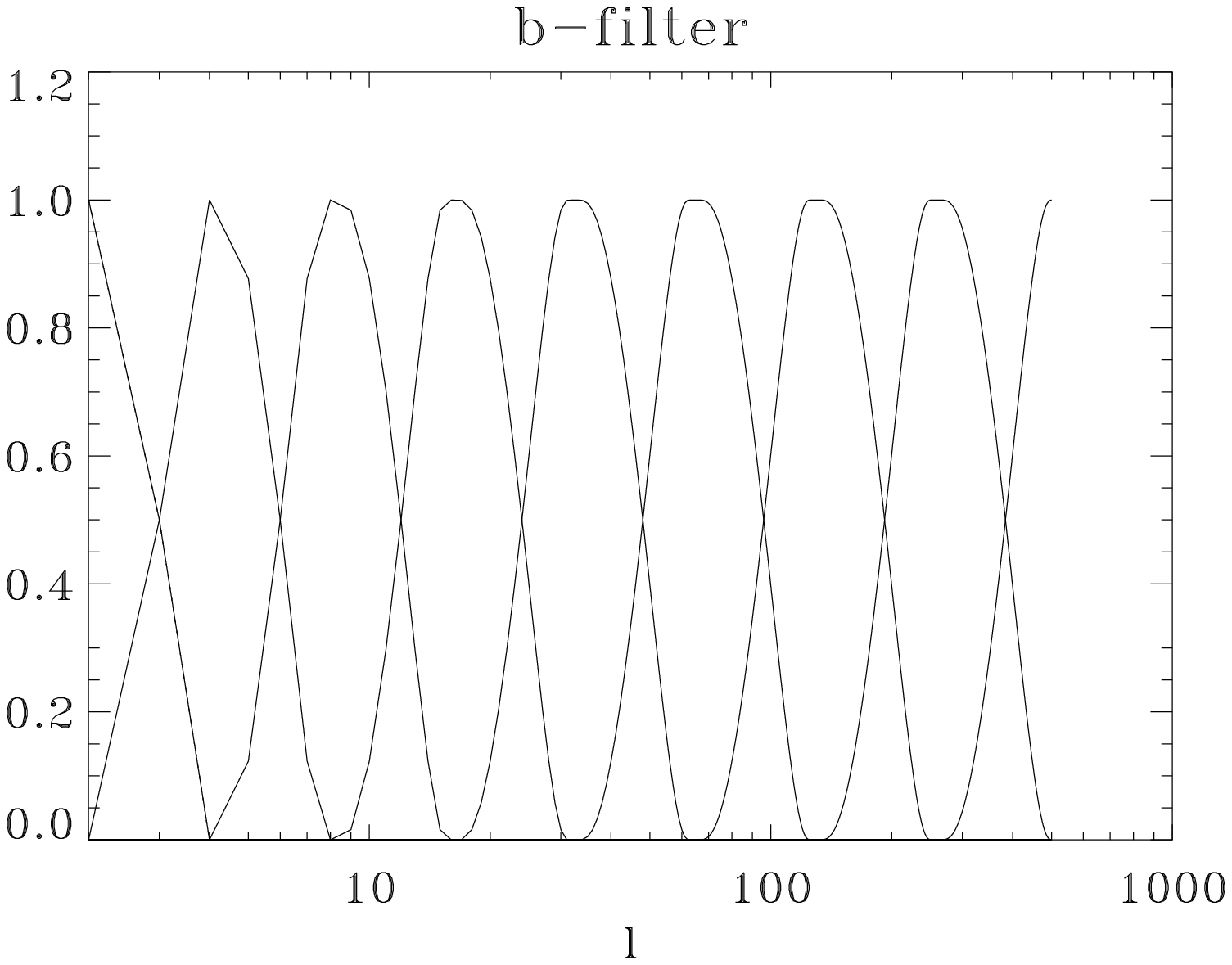}
  \caption{Filter function in $\ell$-space which the needlet construction relies on. Set computed for $B=2$.}
  \label{fig:b_filter}
\efg
For a given field defined on the sphere, $T(\hat\gamma)$, the needlet coefficients are defined as
\bea
  \beta _{jk} &=&\int_{S^{2}}T(\hat\gamma)\psi _{jk}(\hat\gamma)d\Omega  \nonumber \\
             &=&\sqrt{\lambda_{jk}}\sum_{\ell}b\big(\frac{\ell}{B^{j}}\big)\sum_{m=-\ell}^{\ell}a_{\ell m}Y_{\ell m}(\xi _{jk})
  \label{eq:needcof}
\eea
One of the most interesting properties of $\beta_{jk}$ which makes needlets particularly suitable for handling CMB data is the following relation:
\be
  \sum_{jk}\beta_{jk}^2 = \sum_\ell \frac{(2\ell+1)}{4\pi} \mathcal{C}_\ell
\ee 
which describes the square of needlet coefficients as an unbiased
estimator of the angular power spectrum \citep{Pietrobon2006ISW,Marinucci2008NEE}. This means that even if we group multipoles and each needlet peaks at a certain multipole range the total power is conserved: this property is peculiar of the
needlets and it is not shared by other wavelet constructions. Note that this is strictly true only if mask are not applied to the maps, while small differences are expected at low $j$s if sky cuts are present. However when a rather broad and symmetric mask is applied, like in the case of Kq85 WMAP mask, the effect is very small.

The signature of non-Gaussianity appears into the higher moments of a distribution, which are no longer completely specified by the first moment (i.e.~the mean value of the distribution) and the second moment (i.e.~the standard deviation). For a Gaussian distribution, all odd moments are vanishing, while the even ones can be expressed in term of the first two only. We then look for a non-vanishing skewness of the distribution of the needlet coefficients, applying a cubic statistics.

\section{WMAP-5 needlet analysis}
\label{sec:sims_res}

In the following we describe the statistical techniques and  simulations used in order to constrain  $\fnl$.

We started by producing simulations of non-Gaussian CMB maps with the WMAP-5 characteristics, for varying $\fnl$. For each channel [Q1, Q2, V1, V2, W1, W2, W3, W4], we used as input a realization of a simulated primordial non-Gaussian map \citep{Liguori2007NGMaps}; these maps were convolved with the respective beam window functions for each channel 
and a random noise realization was added to each map adopting the  nominal sensitivities and number of hits provided by the WMAP team\footnote{http://lambda.gsfc.nasa.gov/product/map/dr3/m\_products.cfm} \citep{Hinshaw2008WMAP5}. From these single-channel maps we constructed an optimal map via \cite{Jarosik2007}: 
\begin{equation}
T(\gamma) = \sum_{ch} T_{ch}(\gamma) w_{ch}(\gamma)
\end{equation}
where $\gamma$ represents a direction on the sky (which, in practice, is identified with a specific pixel in the Healpix scheme \citep{Gorski2005Healpix}), and
$w_{ch} = n_h(\gamma)/\sigma^2_{ch}/\sum_{ch}w_{ch}$.
where $n_h$ is the number of observations of a given pixel
and $\sigma_{ch}$ the nominal sensitivity of the channel, estimated by WMAP team. We finally applied the WMAP mask Kq85 and degraded the resulting map to the resolution of $N_{\rm }= 256$.  At the end of this procedure we were left with realistic Monte Carlo simulations of the CMB sky as seen from WMAP-5, containing different level of primordial non-Gaussianity parameterised by the value of $\fnl$.

Then, we extracted the needlet coefficients $\beta_{jk}$ from the simulated maps for a given $B$. For each $j$ resolution, the needlet coefficients can be visualised as a sky map, where $k$ is the pixel number. We calculated the skewness of the reconstructed coefficients maps over the unmasked region, as:
\be
  \label{eq:skew}
  S_j = \frac{1}{\tilde{N_p}}\sum_{k^\prime} \frac{(\beta_{jk^\prime}-\langle\beta_{jk^\prime}\rangle)^3}{\sigma_j^3}
\ee
where $\tilde{N_p}$ denotes the number of pixels outside the mask and $\sigma_j$ is the variance of the needlet coefficients at the $j$ resolution. This procedure allows us to build an empirical statistical distribution of the skewness as a function of $\fnl$.
Finally, we calculated the skewness from the real data of the
foreground-reduced WMAP 5-year Q, V and W channels data, using the
same procedure applied to the simulated maps. The comparison of the real data skewness to the simulated distributions allowed us to set limits on the non-Gaussian signal in the data.

A non-vanishing skewness represents a deviation from a Gaussian distribution and could give a fundamental handle on the physics responsible for inflation and the generation of primordial fluctuations. In general we expect the needlet coefficients to pick up signal at different angular scales as a function of both $j$ and $B$, making different sets sensitive to non-Gaussianity in specific ways. This could be indeed a powerful tool when looking for imprint of specific models of non-Gaussianity. This feature is enhanced by the statistics itself we consider in our analysis. Since $S\propto\beta_{jk}^3$ we have an intrinsic modulation in the power of the cube of the needlet
coefficients. Figure \ref{fig:b_sum} shows this effect compared to the square of the $b_\ell$ function.
\bfg
  \incgr[width=\columnwidth]{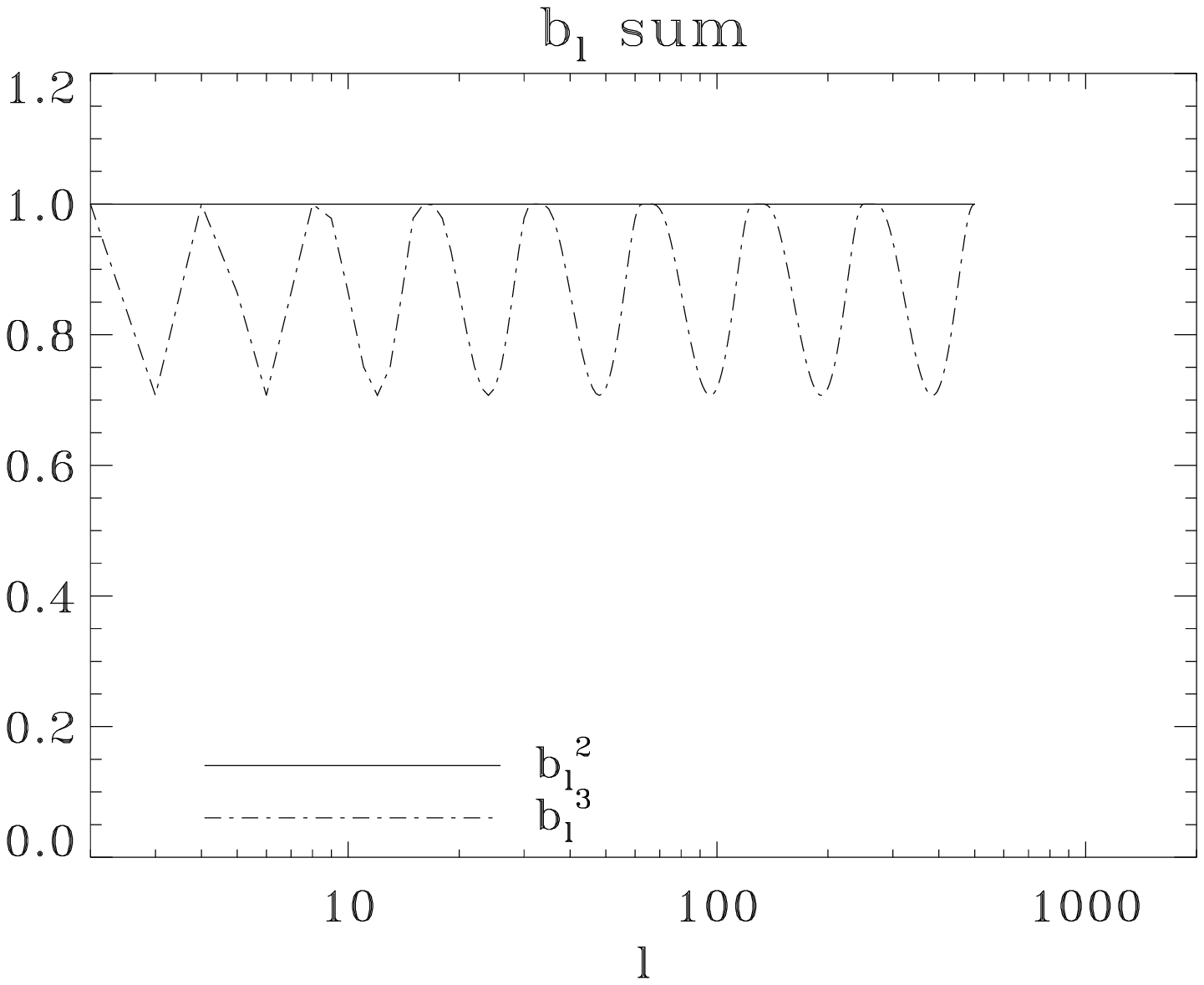}
  \caption{Solid line sum of the $b_\ell^2$; dot-dashed line sum of the $b_\ell^3$. While the former is equal to 1 for the entire range of multipole, the latter is not.}
  \label{fig:b_sum}
\efg

For this reason it does make sense to compute the statistics defined in Eq.~\ref{eq:skew} for several sets of needlets. In particular we employed values of $B$ from $1.8$
to $4.5$, choosing the step in order to span as homogeneously as possible the entire range of multipoles $\ell=2$ to $\ell=500$. The set of $B$ we have considered is [1.8, 1.9, 2.0, 2.15, 2.5, 3.0, 3.5, 4.0, 4.5]. We also tried finer samplings of $B$, but no additional information resulted for the sampling considered.

\begin{figure*}
\incgr[width=.33\textwidth]{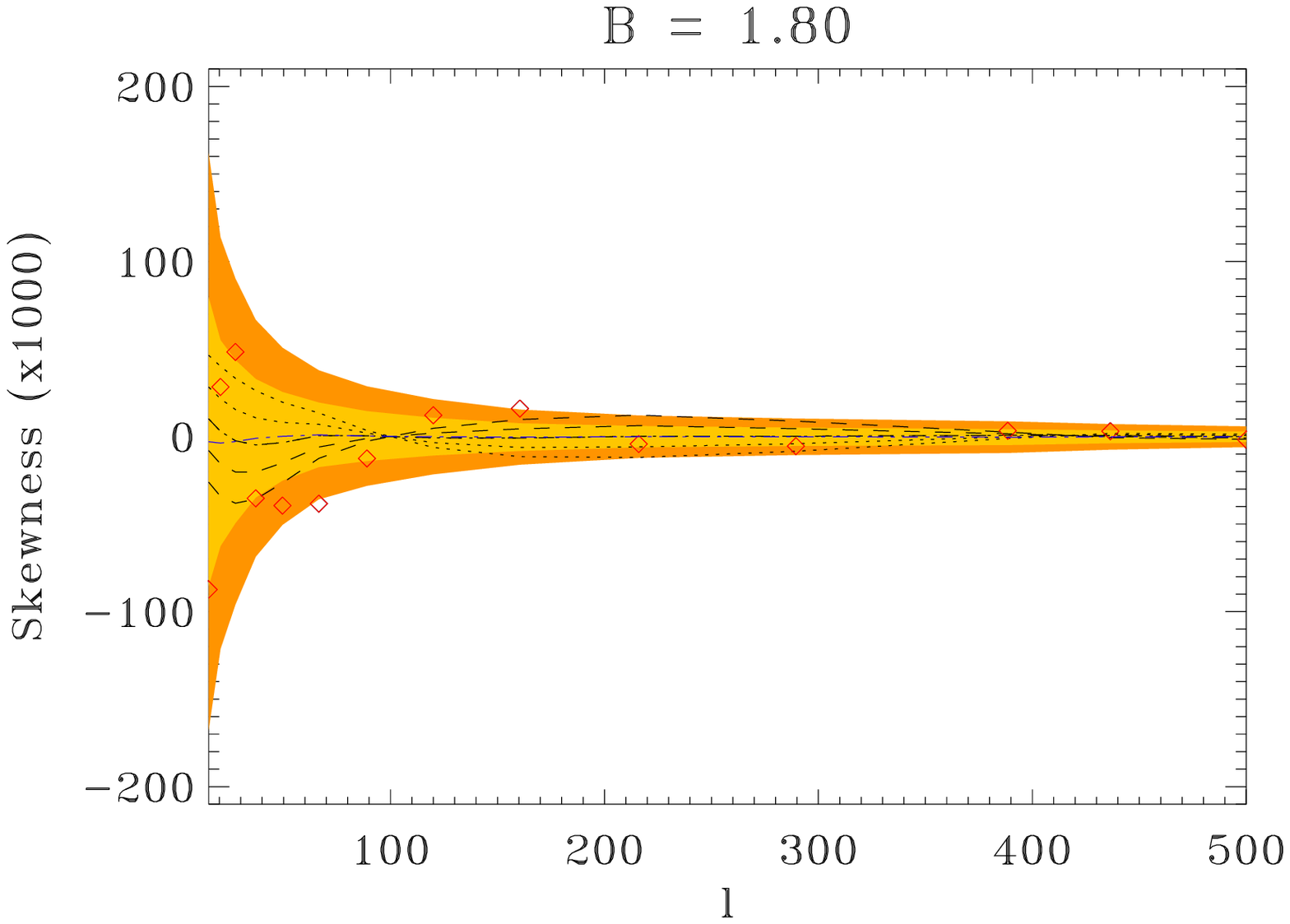}
\incgr[width=.33\textwidth]{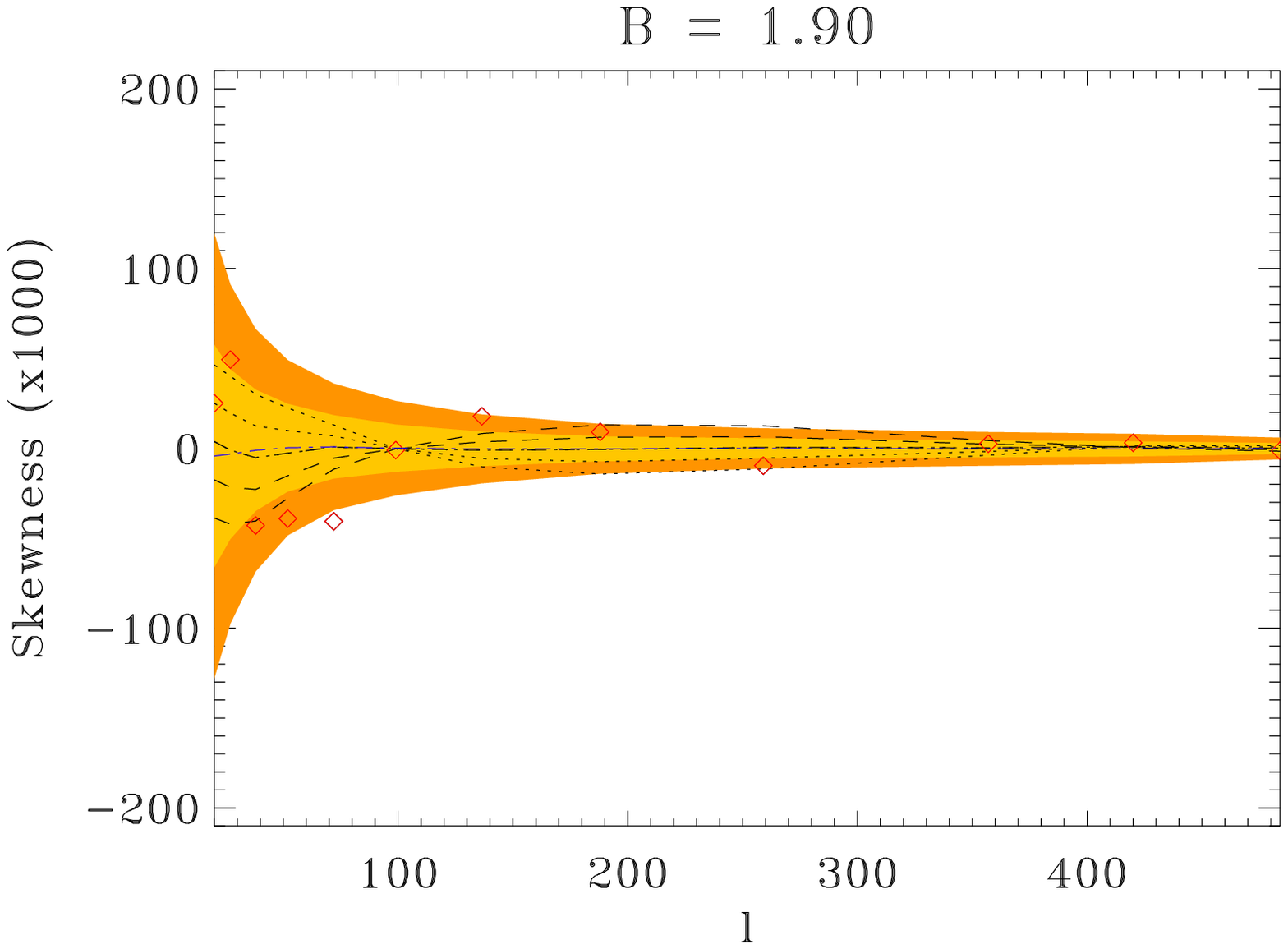}
\incgr[width=.33\textwidth]{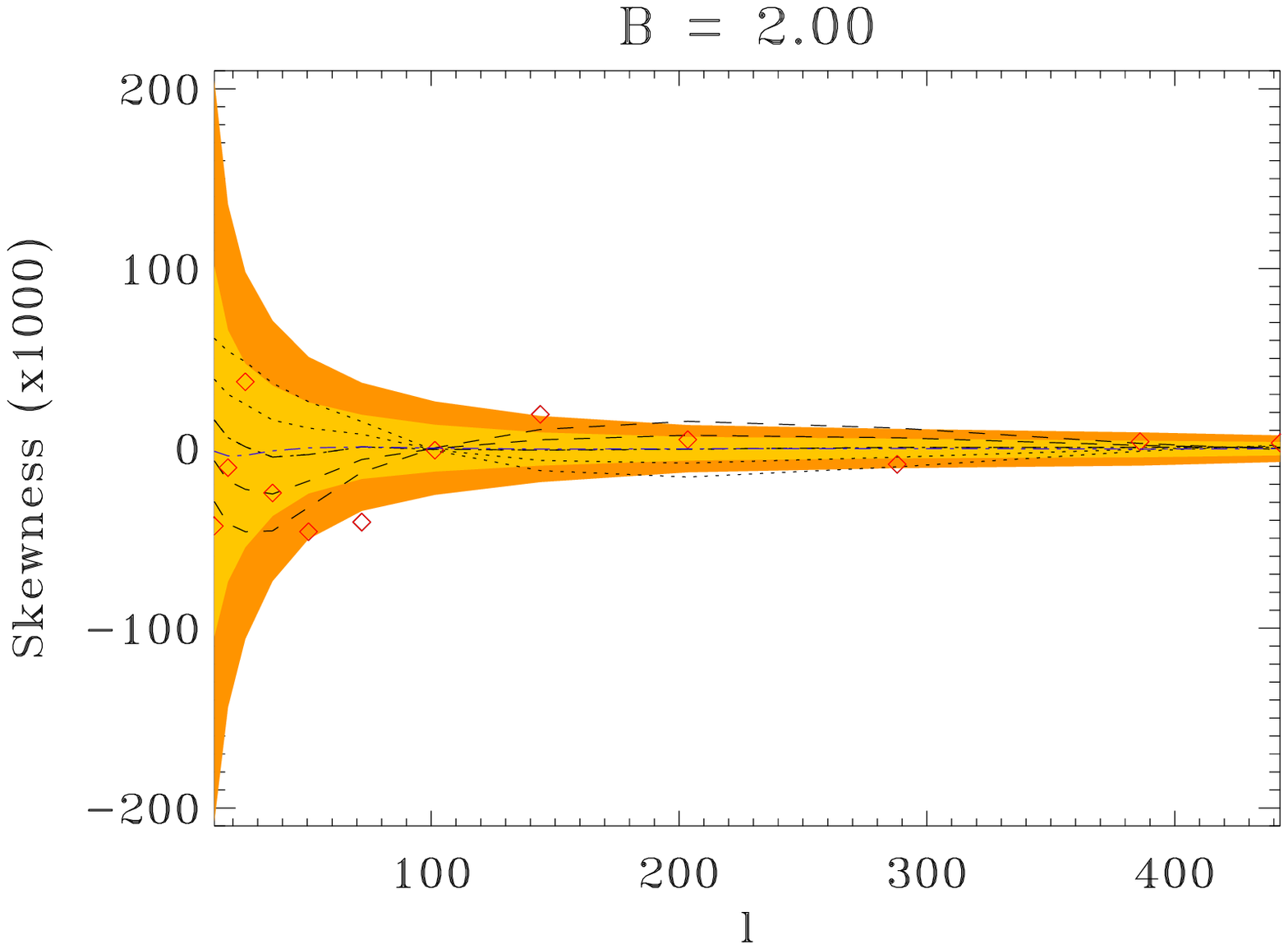}
\incgr[width=.33\textwidth]{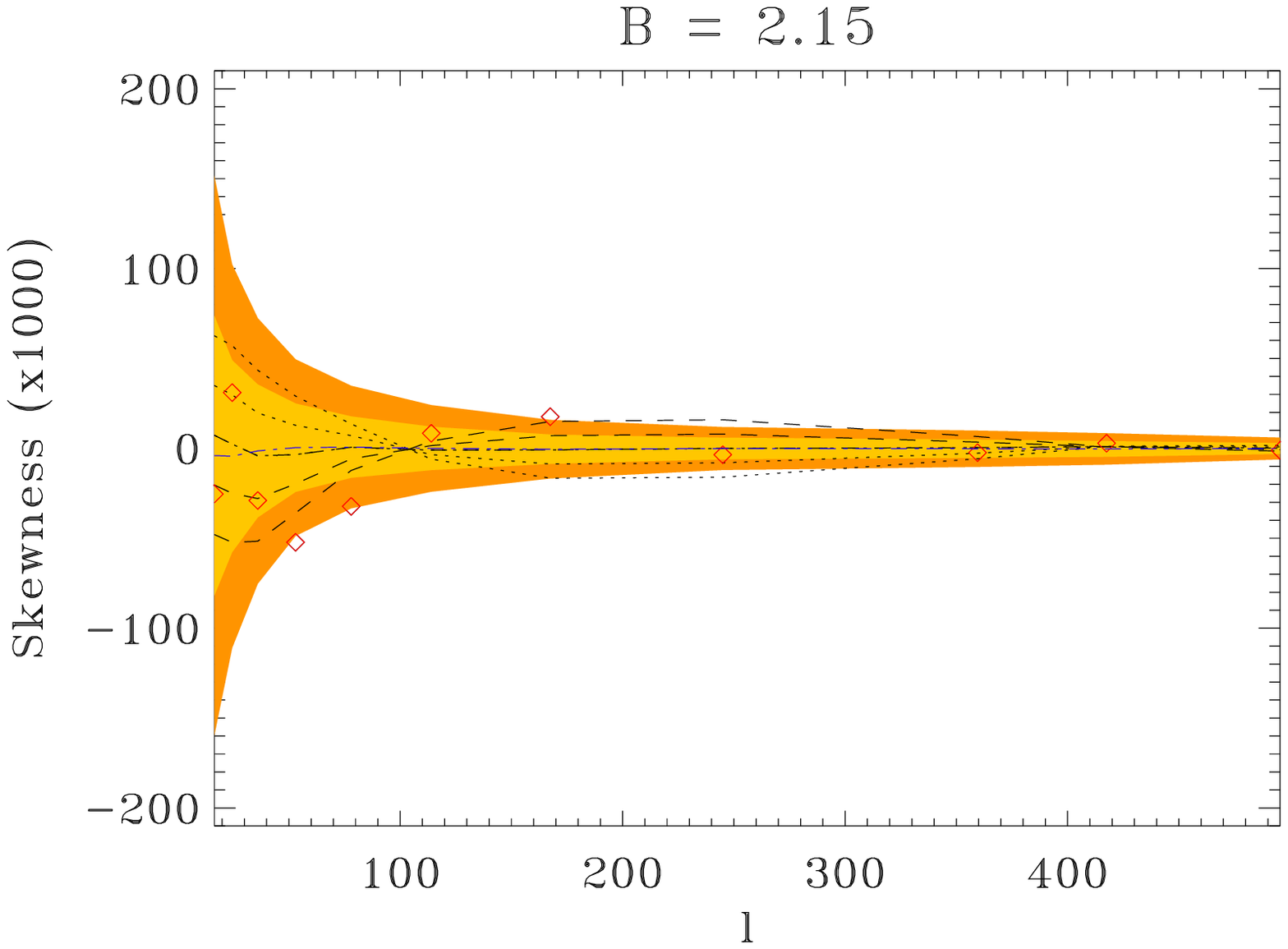}
\incgr[width=.33\textwidth]{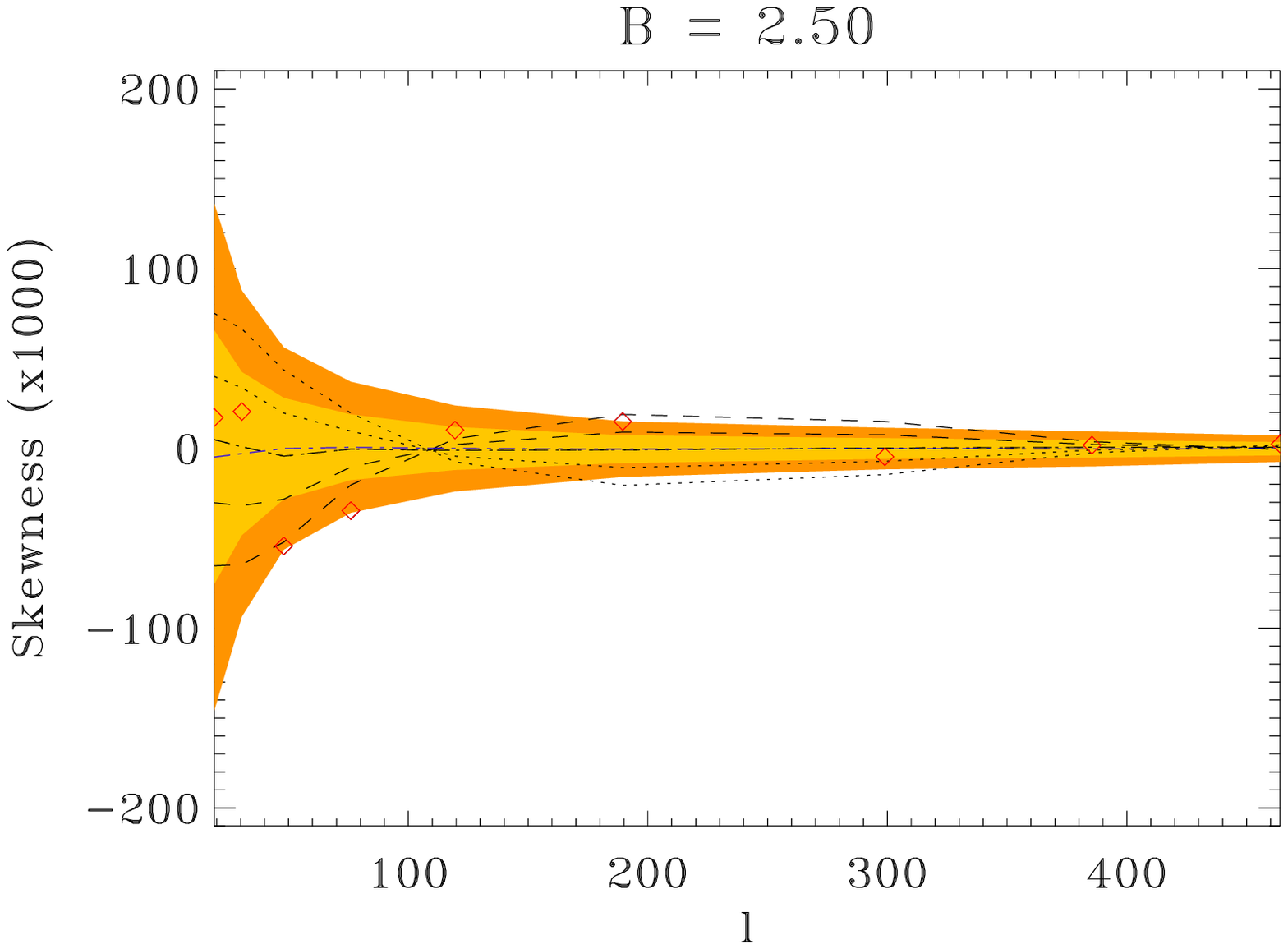}
\incgr[width=.33\textwidth]{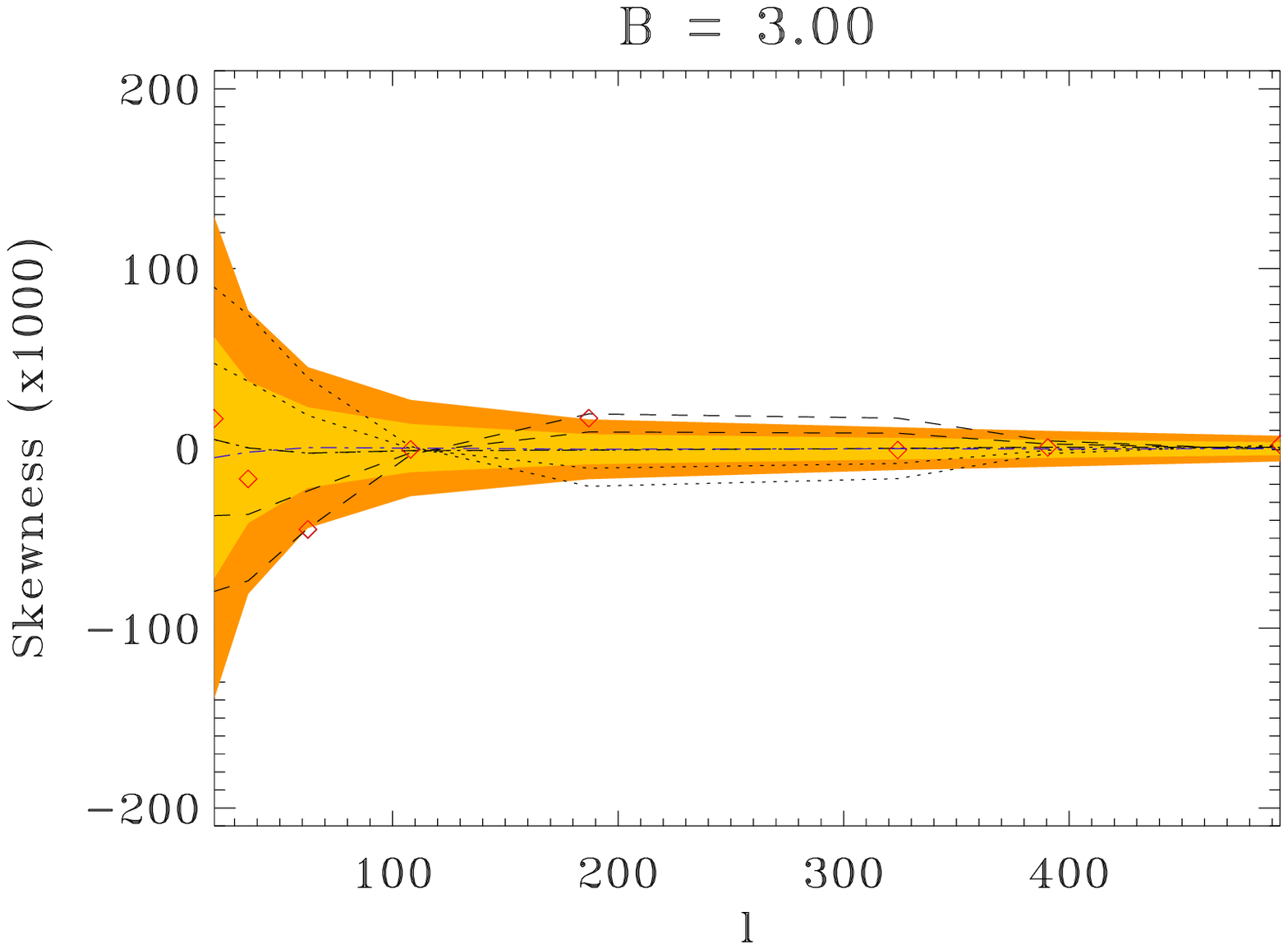}
\incgr[width=.33\textwidth]{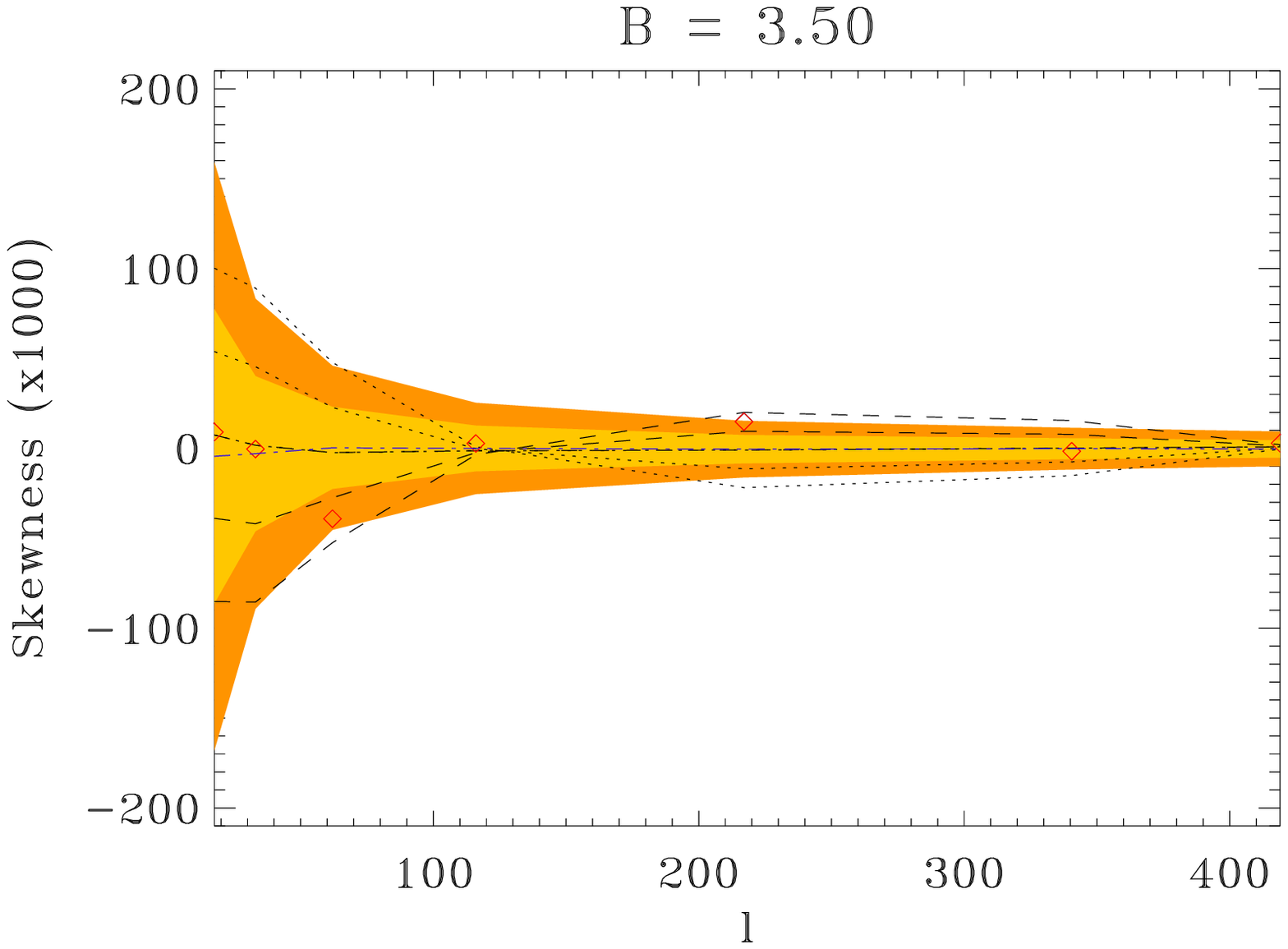}
\incgr[width=.33\textwidth]{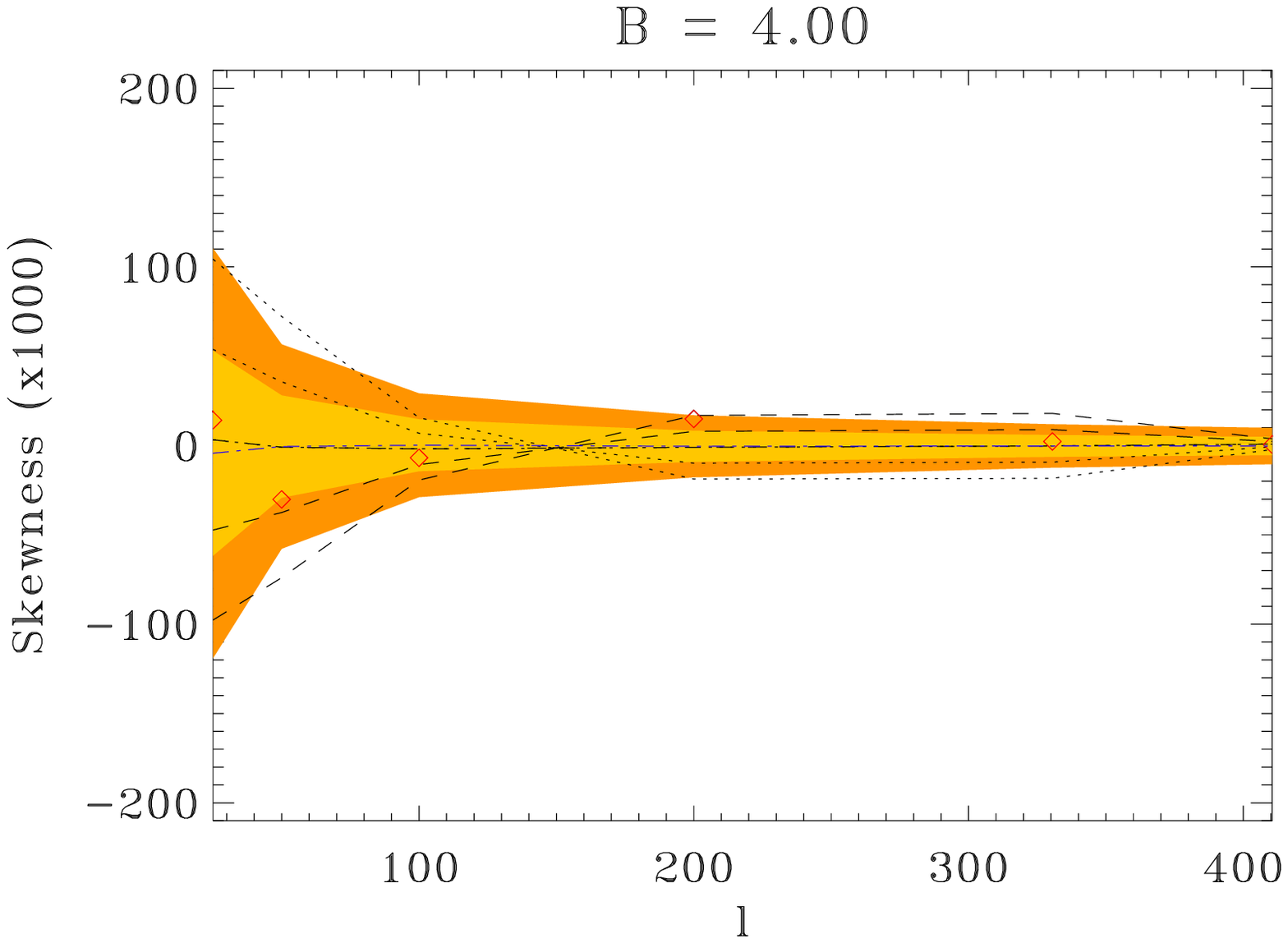}
\incgr[width=.33\textwidth]{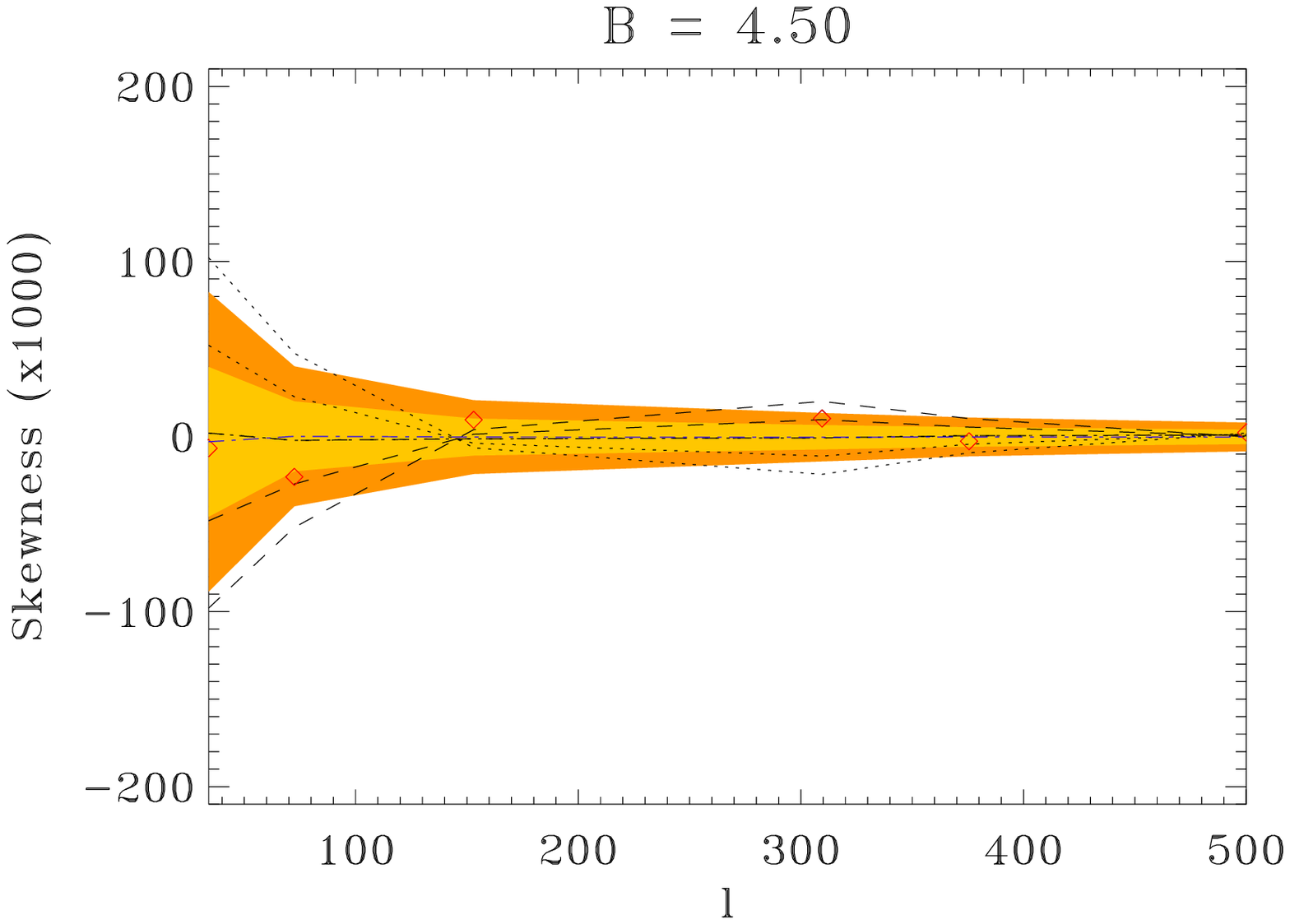}
\caption{Skewness of needlet coefficients for different values of parameter B as a
function of multipole $\ell$ (logarithmic function of the resolution $j$). Shaded areas represents 1 and 2 sigma confidence
levels calculated from 10000 Gaussian Monte Carlo simulations with
beam, noise level, and mask of WMAP 5-year data. Dashed (dotted) lines
correspond to the averages over 100 primordial non-Gaussian maps with
$\fnl=200,400 \; (-200, 400)$. Diamond are the WMAP 5-year
data.}
\label{fig:single_skew}
\end{figure*}

In Fig.~\ref{fig:single_skew} we report the skewness of the needlet coefficients, computed with the experimental set-up described above,  for several values of $B$ as a function of the multipole $\ell$  where the resolution $j$ peaks. The curves deviating from zero corresponds to the effect due to the primordial  non-Gaussianity for positive (dashed lines) and negative (dotted lines) values of $\fnl$, while the yellow and orange bands correspond to the $1\sigma$ and $2\sigma$ levels respectively. Diamonds represent the results of WMAP 5-year data.

\subsection{$\chi^2$ analysis}
In order to estimate $\fnl$ we minimised the quantity:
\be
  \chi^2(\fnl) = (X^d-\langle X(\fnl)\rangle)^T C^{-1}  (X^d-\langle X(\fnl)\rangle).
\ee
Here $X$ is a vector composed by the set of skewness of our set $(B,j)$.
The averages $\langle X(\fnl)\rangle$ were calculated via Monte Carlo simulation over 100 primordial non-Gaussian maps.  
Formally, $C^{-1}$ is dependent on $\fnl$ as well but it has been
shown \citep{SpergelGoldberg1999,KomatsuSpergel2001} that for the mildly level of non-Gaussianity we expect this dependence is weak and can be estimated by Gaussian simulations.  
We found that calculating $C^{-1}$ from 10000 Monte Carlo simulations gives very stable estimate.

\begin{figure*}
\begin{center}
\incgr[width=0.8\textwidth]{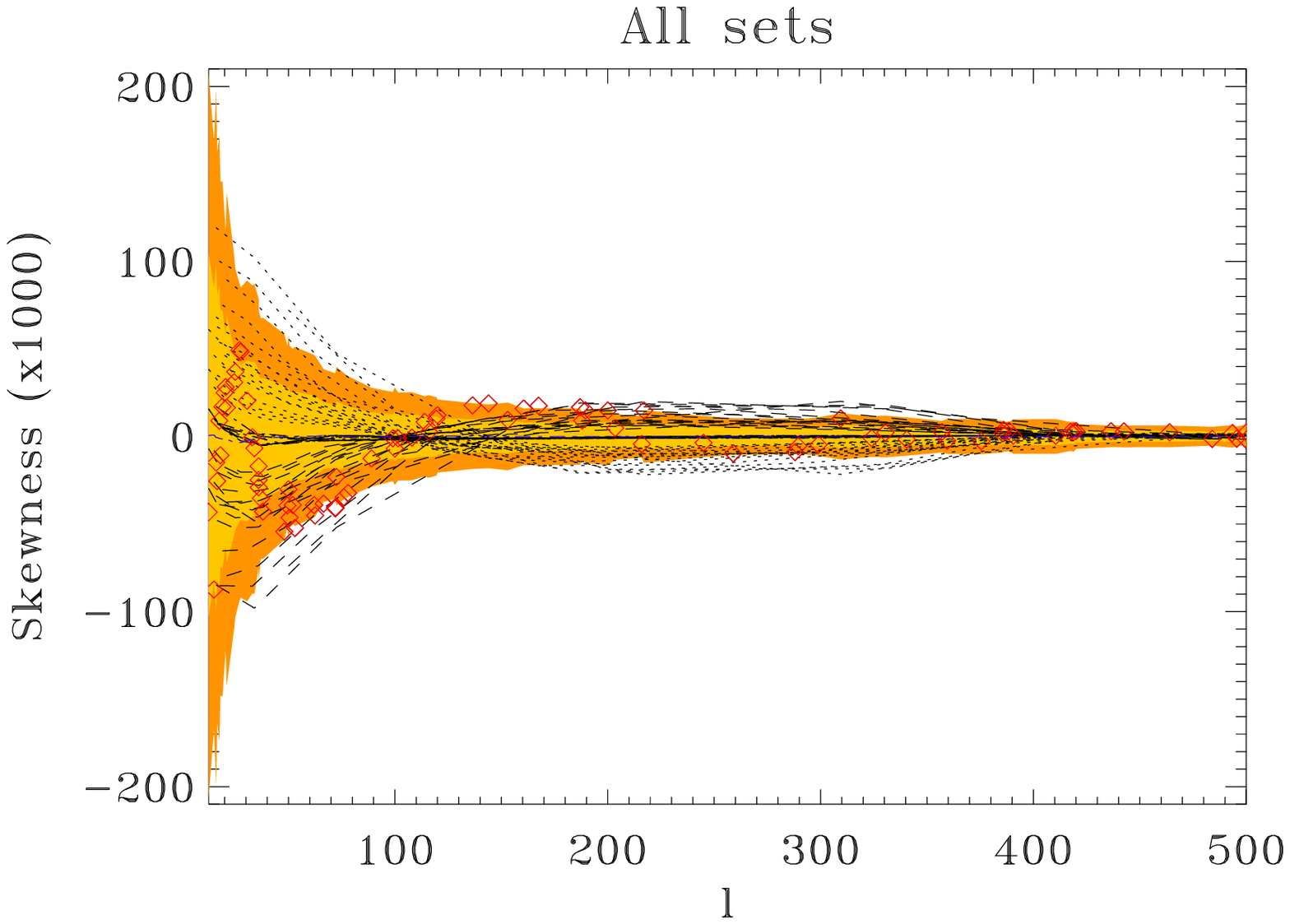}
\caption{Skewness of needlet coefficients for the entire set of parameter B as a
function of multipole $\ell$. Shaded areas represents 1 and 2 $\sigma$ confidence
levels calculated from 10000 Gaussian Monte Carlo simulations with
beam, noise level, and marks of WMAP 5-year data. Dashed (dotted) lines
correspond to the averages over 100 primordial non-Gaussian maps with
$\fnl=200,400 \; (-200, 400)$. Diamond are the WMAP 5-year
data.}
\label{fig:global_skew}
\end{center}
\end{figure*}

The result is shown in Fig.~\ref{fig:global_skew}: $\fnl$ is
estimated to be $20$ with  $-30<\fnl<70$ and $-80 <\fnl<120 $ at 1 and $2\sigma$ respectively.
These results show no significant deviation from the Gaussian
hypothesis. This is not in contrast with the values found by
\cite{YadavWandelt2007}, since we have performed our analysis on maps
with the maximum multipole corresponding to $\ell_{\rm max}=500$, whereas \cite{YadavWandelt2007} clearly showed that their results crucially depend on the maximum multipole considered.
The reduced $\chi^2$ of WMAP data is $1.53$ with 85 degrees of freedom. 

As a further consistency check, we performed a goodness-of-fit test by calculating the quantile of the WMAP data from the non-Gaussian
realizations with $\fnl=20$. We found that 21\% of the simulations had a larger $\chi^2$ of the WMAP $\chi^2$, confirming that 
the specifications of our Monte Carlo simulations well describe the
experimental setting of WMAP 5-year data.

\bfg
\incgr[width=\columnwidth]{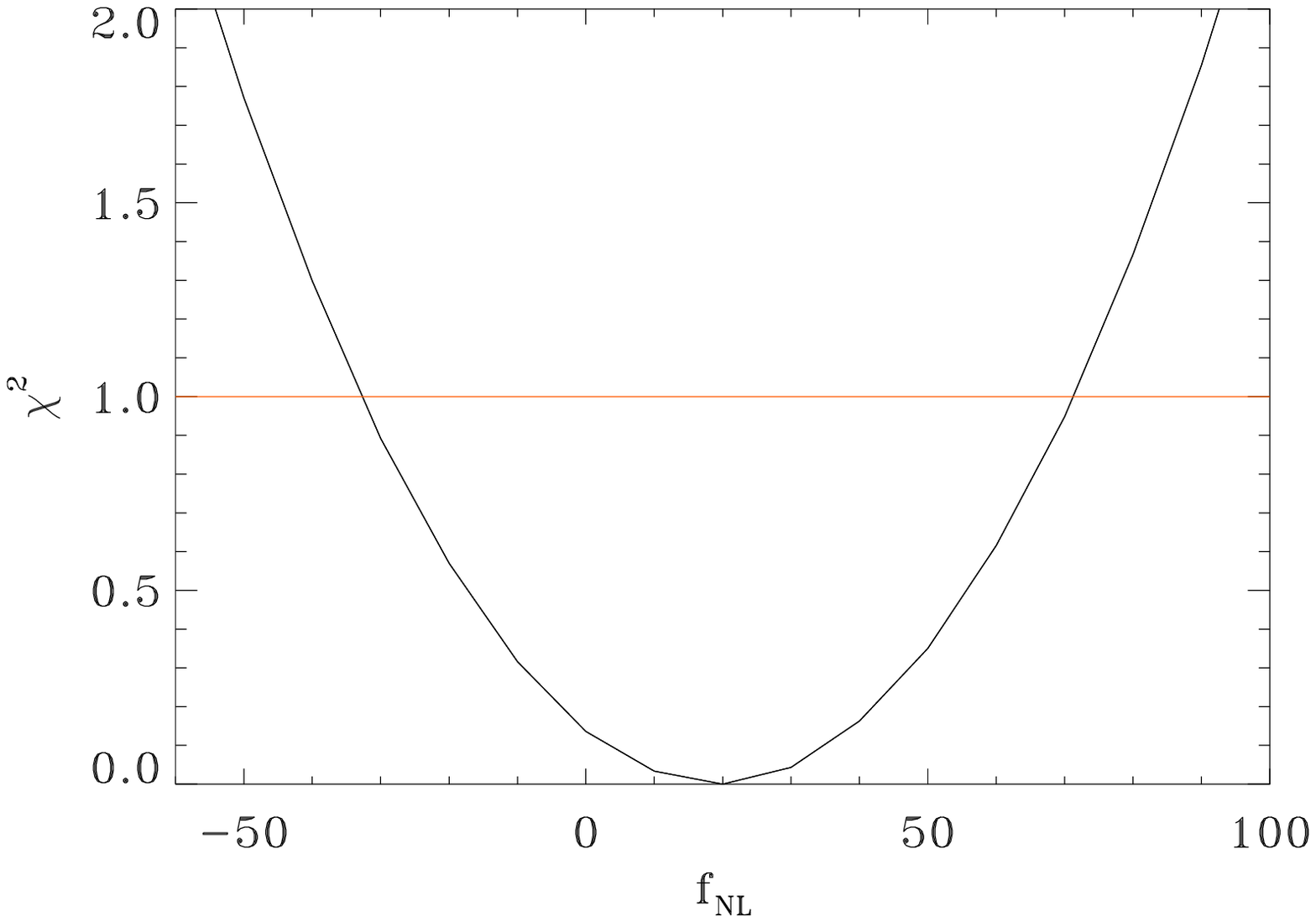}
\caption{The $\Delta\chi^2$ of WMAP 5-year data as a function of $\fnl$.
$\fnl$ is estimated to be $\fnl=20\pm50 $ and  $\fnl=20\pm100 $ at 1
$\sigma$ and 2 $\sigma$ level respectively.}
\label{fig:gchi2_highNGnorm15}
\efg

\subsection{A skewness based $\fnl$ estimator}
Since the primordial non-Gaussianity is a second order effect, it contributes linearly to the skewness through $\fnl$. This means that it is possible to compute from the non-Gaussian simulations the skewness $S(j)$ for $\fnl=1$ and use it as template to build a filter-matching estimator of the non-linear coupling parameter. Assuming that $S_j^{\rm obs} = \fnl\; S_j^{\rm th}|_{\fnl=1}$, where ``{\rm th}'' means the average over the non-Gaussian simulations, we obtain
\be
   \fnl = \frac{\sum_{jj^\prime}S^{\rm obs}_j\mbf{Cov}^{-1}_{jj^\prime}S^{\rm th}_{j^\prime}}{\sum_{jj^\prime}S^{\rm th}_j\mbf{Cov}^{-1}_{jj^\prime}S^{\rm th}_{j^\prime}}
   \label{eq:fnl_estimator}
\ee
where we dropped the subscript $\fnl=1$.
The theoretical skewness computed for $\fnl=1$ is shown in Fig.~\ref{fig:skew_fnl1}.
\bfg
   \incgr[width=\columnwidth]{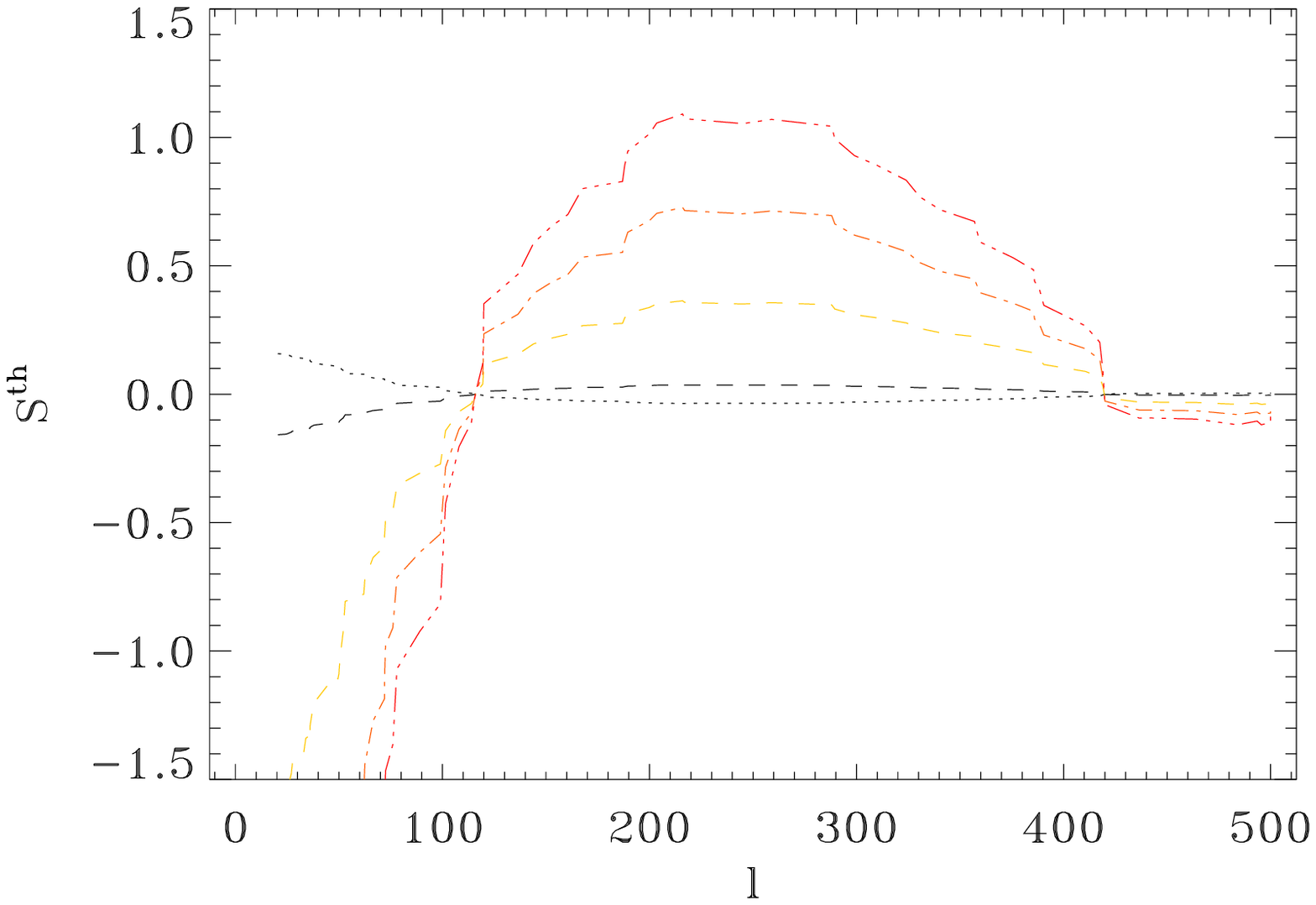}
   \caption{Skewness for $\fnl=\pm1$, respectively dashed and dotted
line, derived from non-Gaussian simulations. Theoretical curves for
$\fnl=10,20,30$ (from bottom to top) are shown too.}
   \label{fig:skew_fnl1}
\efg
We checked that the pipeline applied to simulated non-Gaussian CMB maps does not affect the linear relation: in particular we verified that the average signal we obtain for a given $\fnl$ scales linearly with $\fnl$ itself, meaning for example that we can mimic the signal for $\fnl=\pm400$ by taking the double of that for $\fnl=\pm200$.

The main contribution to the covariance matrix $\mbf{Cov}_{jj^\prime}$ comes from the Gaussian part of gravitational potential; this allows us to estimate the covariance from random Gaussian simulations. According to this assumption, we estimate the error bars on the primordial non-linear coupling parameter computing the standard deviation of the 10000 $\fnl$ estimates resulting from a fresh set of Gaussian simulations, via Eq.~\ref{eq:fnl_estimator}. We find $\fnl=21\pm54$ at 1 sigma confidence level, which is fully consistent with what we found applying the $\chi^2$ statistic. This corroborates the robustness of our procedure and confirms needlets as a suitable tool to study primordial non-Gaussianity.

\vspace{0.25cm}
Our limits on primordial non-Gaussianity are slightly larger than those achieved by 
\cite{Curto2008waveNG} being our $1\sigma$ confidence level ($-30<\fnl<70$) slightly broader
than $-8<\fnl<-111$ at $95\%$ confidence level. However we were limited in our analysis to
multipoles lower than $\ell_{max}=500$, while the strongest constraints
on primordial non-Gaussianity make use of higher angular scales.
       
Our limits on $\fnl$ are quite promising for future experiments such as Planck\footnote{http://www.rssd.esa.int/SA/PLANCK/docs/Bluebook-ESA-SCI(2005)1.pdf}, where sensitivity and angular resolution will be enormously improved.
 
\subsection{Further analysis: binned bispectrum}
\label{subsec:bisBin}

The analysis we performed in the previous sections, based on the
skewness of the needlets coefficients $\beta_{jk}$, is mainly sensitive
to the equilateral configurations, since it is proportional to the
primordial bispectrum computed for $\ell_1\simeq\ell_2\simeq\ell_3$ summed
over all the multipoles up to $\ell_{\rm max}=500$. This can be easily
understood by direct inspection of Eq.~\ref{eq:skew}:
\bea
\label{eq:skew_exp}
S_j &\simeq& \frac{1}{\tilde N_{\rm p}}\sum_k\frac{\beta_{jk}^3}{\sigma_j} \nonumber \\
    &\propto & \sum_{\ell_1 m_1} \sum_{\ell_2 m_2} \sum_{\ell_3 m_3}
b_{\ell_1}^{(j)} b_{\ell_2}^{(j)} b_{\ell_3}^{(j)} a_{\ell_1 m_1} a_{\ell_2 m_2} a_{\ell_3 m_3} \nonumber \\
&& \sum_k Y_{\ell_1 m_1}(\xi_{jk}) Y_{\ell_2 m_2}(\xi_{jk}) Y_{\ell_3 m_3}(\xi_{jk})
\eea
The sum over pixels approximates Gaunt's integral
\bea
&&\sum_k Y_{\ell_1 m_1}(\xi_{jk}) Y_{\ell_2 m_2}(\xi_{jk}) Y_{\ell_3
m_3}(\xi_{jk}) \nonumber \\
&\approx& \int_S\de\Omega Y_{\ell_1 m_1}(\hat\gamma)
Y_{\ell_2 m_2}(\hat\gamma) Y_{\ell_3 m_3}(\hat\gamma) \nonumber \\
&=& \sqrt{\frac{(2\ell_1+1)(2\ell_2+1)(2\ell_3+1)}{4\pi}} \nonumber \\
&&\left(
\begin{array}{ccc}
\ell_1 & \ell_2 & \ell_3 \\
0 & 0 & 0 \\
\end{array}
\right) \nonumber \\
&&\left(
\begin{array}{ccc}
\ell_1 & \ell_2 & \ell_3 \\
m_1 & m_2 & m_3 \\
\end{array}
\right)
\eea
and introducing the estimated bispectrum as the average over $m_1$,
$m_2$ and $m_3$
\bea
\hat{\rm B}_{\ell_1\ell_2\ell_3} &\equiv& \langle
a_{\ell_1m_1}a_{\ell_2m_2}a_{\ell_3m_3}\rangle \nonumber \\
&=& \sum_{m_1m_2m_3}\left(
\begin{array}{ccc}
\ell_1 & \ell_2 & \ell_3 \\
m_1 & m_2 & m_3 \\
\end{array}
\right)a_{\ell_1m_1}a_{\ell_2m_2}a_{\ell_3m_3} \nonumber
\eea
 we obtain the following relation:
\bea
S_j &\propto& \sum_{\ell_1\ell_2\ell_3}b_{\ell_1}^{(j)}b_{\ell_2}^{(j)}b_{\ell_3}^{(j)}\sqrt{\frac{(2\ell_1+1)(2\ell_2+1)(2\ell_3+1)}{4\pi}} \nonumber \\
&\times&\left(
\begin{array}{ccc}
\ell_1 & \ell_2 & \ell_3 \\
0 & 0 & 0 \\
\end{array}
\right)\hat{\rm B}_{\ell_1\ell_2\ell_3}
\eea
Since the filter functions $b_{\ell_i}^{(j)}$ are computed for the
same resolution $j$, the biggest contribution to the bispectrum comes
from the equilateral configurations, plus a smaller effect of
non-equilateral triangles when, for the same $j$,
$\ell_1\neq\ell_2\neq\ell_3$.

A qualitative improvement in constraining the parameter $\fnl$ can be
achieved by adding in the estimator the effect of the squeezed configurations,
considering the product of three $\beta_{jk}$ with $j_1\neq j_2\neq
j_3$. The skewness of needlet coefficients can be generalised into
\bea
S_{j_1j_2j_3} &=&\frac{1}{\tilde{N}_p}
\sum_k\frac{\beta_{j_1k}\beta_{j_2k}\beta_{j_3k}}{\sigma_{j_1}\sigma_{j_2}\sigma_{j_3}}
\\
&\propto& \sum_{\ell_1\ell_2\ell_3}b_{\ell_1}^{(j_1)}b_{\ell_2}^{(j_2)}b_{\ell_3}^{(j_3)}\sqrt{\frac{(2\ell_1+1)(2\ell_2+1)(2\ell_3+1)}{4\pi}} \nonumber \\
&\times&\left(
\begin{array}{ccc}
\ell_1 & \ell_2 & \ell_3 \\
0 & 0 & 0 \\
\end{array}
\right)\hat{\rm B}_{\ell_1\ell_2\ell_3} \nonumber
\eea
$S_{j_1j_2j_3}$ can be seen as a \emph{binned bispectrum}, a smooth and
combined component of the angular bispectrum.

We repeated the needlet analysis applying this new estimator to the
same set of WMAP 5-year data and simulations for the choice of the
needlet parameter $B=3.5$, which has the highest
signal-to-noise ratio among the set chosen in the previous analysis. The minimization of
the $\chi^2$ gives $\fnl=30\pm40$, which is consistent with what we
found applying $S_j$. The $\chi^2$ for WMAP 5-year data is shown in Fig.~\ref{fig:binBis_chi2}
\bfg
\incgr[width=\columnwidth]{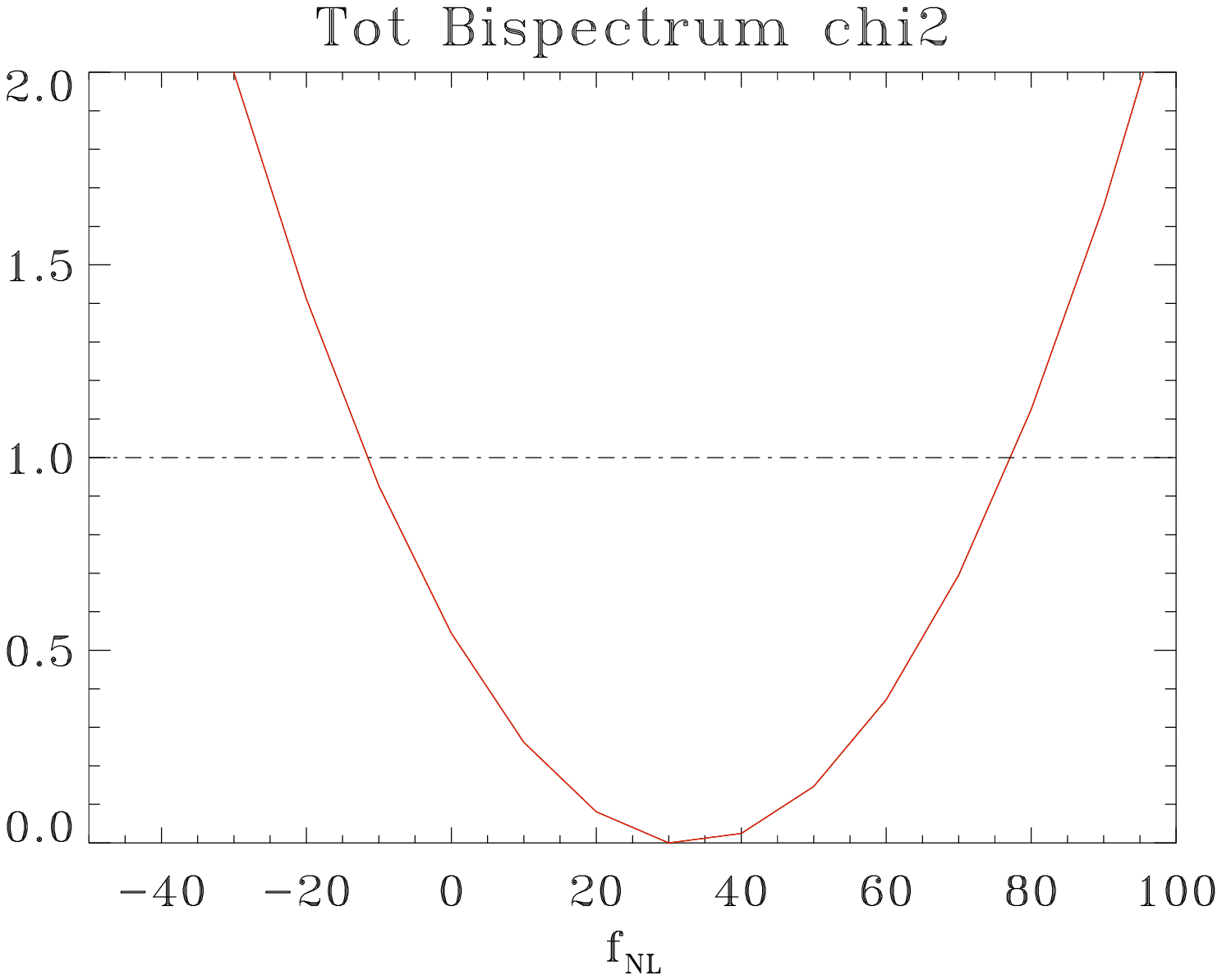}
\caption{The $\Delta\chi^2$ of WMAP 5-year data as a function of
$\fnl$ computed from the binned bispectrum. $B$ is chosen $3.50$ which
had the highest signal-to-noise ratio among those chosen in the
previous analysis. $\fnl$ is estimated to be $\fnl=30\pm40 $ and  $\fnl=30\pm80 $ at 1
$\sigma$ and 2 $\sigma$ level respectively.}
\label{fig:binBis_chi2}
\efg

Recently an analysis based on a cubic estimator analogous to 
$S_{j_1j_2j_3}$ has been performed by \cite{Curto2008waveNG,Curto2009}
who include the effect of squeezed configurations in the Spherical Mexican
Hat Wavelet; and by \cite{Rudjord2009needBis} using
a set of needlets characterised by a different $B$ parameter obtaining $\fnl=84\pm40$. While
the difference in the value of $\fnl$ can be due to the higher number
of multipoles considered in the analysis, $\ell_{\rm max}=1000$, and
it is consistent with the result of \cite{YadavWandelt2007}, it is important
that the estimated error bars are fully consistent with
ours.
\section{Conclusions}
\label{sec:concl}

Primordial non-Gaussianity is becoming one of the keys to understand the
physics of
the early Universe. Several tests have been developed and applied to WMAP
data to
constrain the non-linear coupling parameter $\fnl$.
Recently, different methods \citep{YadavWandelt2007,Curto2008waveNG}
brought to different constraints on $\fnl$ using similar datasets, WMAP 3-year
and WMAP 5-year respectively. The two approaches have been shown to have the
same power in constraining primordial non-Gaussianity, while they
obtained different best fit values for $\fnl$. Whereas this might be due to
the masks applied to the datasets, it certainly underlines
the complexity and difficulty of measuring $\fnl$.
The next generation of experiments will provide data with excellent angular
resolution and signal-to-noise ratio which will be decisive to
confirm or confute the measurements of  $\fnl$ of the references above.
In this respect, it will be even important to constrain $\fnl$ with
different methods in order to get a more robust detection or to spot
spurious presences of non-Gaussian signal. Moreover, integrated
estimators, not based on Wiener filters, are differently sensitive to
the non-linear coupling and can be useful to address exotic
non-Gaussian models which predict high values of $\fnl$ and whose
bispectrum evaluation, for instance, may require prohibitive computational time due to the convolution in the bispectrum formula.

In this work, we constrained the primordial non-Gaussianity parameter
$\fnl$ by developing the needlets formalism and applying it to the
WMAP 5-year CMB data. We estimated $\fnl$ to be $20$ with
$-30<\fnl<70$ and $-80<\fnl<120$ at 1 and 2 sigma respectively, then
consistent with the Gaussian hypothesis. We performed two different
analyses, the $\chi^2$ statistics and an estimator based on the
skewness of the primordial non-Gaussian sky, finding an excellent
agreement between the two results. Needlets have been proven to be a
well understood tool for CMB data analysis, sensitive to the
primordial non-Gaussianity. Since the skewness of the
needlets coefficients is mainly sensitive to the equilateral
triangle configurations, we improved our estimator computing the three point
correlation function in needlet space which indeed recovers the
signal due to squeezed triangle configurations. We obtain
$\fnl=30\pm40$ at $68\%$ confidence level, consistent with the
previous analysis.
Our constraints are slightly broader
than those achieved by \cite{Curto2008waveNG} and not in contrast with the values found by \cite{YadavWandelt2007} 
since we were limited by a smaller range of multipoles,
whereas the tighter constraints on $\fnl$ crucially depend on the maximum multipole considered.
       
Our limits on $\fnl$ are quite promising for future experiments like Planck, whose sensitivity and angular resolution will be enormously improved.

\section*{Acknowledgements}
We thank Frode K.~Hansen, Michele Liguori and Sabino Matarrese for
providing us with the primordial non-Gaussian map dataset. We are
grateful to Robert Crittenden for useful discussions. We thank
Domenico Marinucci for interesting discussions. D.~P.~ thanks Asantha Cooray and the UCI Physical Sciences Department where this work has been partially performed.

\bibliography{Bng_need}

\end{document}